\documentclass[twocolumn,showpacs,preprintnumbers,amsmath,amssymb]{revtex4}

\usepackage{graphicx}
\usepackage{dcolumn}
\usepackage{bm}

\newcommand{\Mpcinv}{\mbox{ Mpc$^{-1}$}}

\newcommand{\keV}{\mbox{ keV}}
\newcommand{\GeV}{\mbox{ GeV}}
\newcommand{\eV}{\mbox{ eV}}
\newcommand{\kel}{\mbox{ K}}
\newcommand{\mkel}{\mbox{ mK}}

\newcommand{\yr}{\mbox{ yr}}
\newcommand{\hr}{\mbox{ hr}}

\newcommand{\secinv}{\mbox{ s$^{-1}$}}
\newcommand{\MHz}{\mbox{ MHz}}

\newcommand{\hunits}{\mbox{ km s$^{-1}$ Mpc$^{-1}$}}

\newcommand{\recunits}{\mbox{ cm$^{3}$ s$^{-1}$}}

\newcommand{\bxhi}{\bar{x}_{\rm HI}}
\newcommand{\xhi}{x_{\rm HI}}

\newcommand{\bxion}{\bar{x}_i}
\newcommand{\xion}{x_i}
\newcommand{\dtb}{\delta T_b}
\newcommand{\bdtb}{\bar{\delta T}_b}

\newcommand{\lya}{Ly$\alpha$ }

\newcommand{\bk}{{\bf k}}

\newcommand{\ga}{{\ \gtrsim \ }}
\newcommand{\la}{{\ \lesssim \ }}
\newcommand{\deriv}{{\rm d}}
\def\VEV#1{\left\langle #1\right\rangle} 

\newcommand{\apjl}{Astrophys. J. \ }

\newcommand{\aap}{Astron. Astrophys. \ }
\newcommand{\aj}{Astron. J. \ }
\newcommand{\mnras}{Mon. Not. R. Astron. Soc. \ }

\newcommand{\physrep}{Phys. Rep. \ }

\begin{document}

\title{The Effects of Dark Matter Decay and Annihilation on the  \\ High-Redshift 21 cm Background}

\author{Steven R. Furlanetto}
\email{steven.furlanetto@yale.edu}
\affiliation{Yale Center for Astronomy and Astrophysics, Yale University, 260 Whitney Avenue, New Haven, CT 06511}

\author{S. Peng Oh}
\affiliation{Department of Physics, University of California, Santa Barbara, CA 93106}

\author{Elena Pierpaoli}
\affiliation{Physics and Astronomy Department, University of Southern California, Los Angeles, CA 90089-0484, USA}
\affiliation{Theoretical Astrophysics, California Institute of Technology, Mail Code 130-33, Pasadena, CA 91125}

\date{\today}

\begin{abstract}

The radiation background produced by the 21 cm spin-flip transition of neutral hydrogen at high redshifts
can be a pristine probe of fundamental physics and cosmology.  At $z \sim 30$--$300$, the intergalactic medium (IGM) is visible in 21 cm absorption against the cosmic microwave background (CMB), with a strength that depends on the thermal (and ionization) history of the IGM.  Here we examine the constraints this background can place on dark matter decay and annihilation, which could heat and ionize the IGM through the production of high-energy particles.  Using a simple model for dark matter decay, we show that, if the decay energy is immediately injected into the IGM, the 21 cm background can detect energy injection rates $\ga 10^{-24}$ eV cm$^{-3}$ sec$^{-1}$.  If all the dark matter is subject to decay, this allows us to constrain dark matter lifetimes $\la 10^{27} \sec$. Such energy injection rates  are much smaller than those typically probed by the CMB power spectra.  The expected brightness temperature fluctuations at $z\sim 50$ are a fraction of a mK and can vary from the standard calculation by up to an order of magnitude, although the difference can be significantly smaller if some of the decay products free stream to lower redshifts.  For self-annihilating dark matter, the fluctuation amplitude can differ by a factor $\la 2$ from the standard calculation at $z \sim 50$.  Note also that, in contrast to the CMB, the 21 cm probe is sensitive to both the ionization fraction and the IGM temperature, in principle allowing better constraints on the decay process and heating history.  We also show that strong IGM heating and ionization can lead to an enhanced H$_2$ abundance, which may affect the earliest generations of stars and galaxies.

\end{abstract}


\maketitle

\section{\label{intro}Introduction}

The cosmic ``dark ages," stretching from the last scattering surface of the cosmic microwave background (CMB) at $z \sim 1100$ to the formation of the first luminous sources at $z \sim 30$, are one of the last frontiers for observational cosmology.  The problem is the difficulty of finding any probes:  because the intergalactic medium (IGM) has mostly decoupled from the CMB, there is no background radiation field against which we can study it, and (by definition) there are no local light sources.  

This is unfortunate because, at least in principle, the physics of the dark ages is sufficiently simple that we can hope to understand it in detail.  The only factors that enter are the CMB, the expanding Universe, recombinations, Compton scattering (which couples the CMB to the IGM), and gravitational growth -- almost entirely in the linear regime.  Thus, the IGM properties during the dark ages (and especially its fluctuations) constitute a strong test of fundamental cosmological parameters in an entirely analogous way to the CMB.  Conversely, if we take the physics as well-understood, probes of the dark ages would offer stringent tests of exotic physics.

Two examples are dark matter decay and annihilation.  If either of these processes happen (even for only a fraction of the dark matter), they could inject high-energy photons (or other particles) into the IGM.  These would then scatter and deposit some or all of their energy as ionizations and heat.  An altered reionization history may affect the total optical depth and, as a consequence, the CMB power spectrum.
 Decaying dark matter particles were in fact  initially advocated  as a possible explanation for the high optical depth  detected in the first year of \emph{WMAP} data \cite{chen04-decay, hansen04, kasuya04, avelino04, pierpaoli04,Bean03}.  Such data allowed an upper limit on the energy injection  rate and consequently on the decay time ($t \ge H_0^{-1}$).  (These constraints, and others mentioned throughout this paper, were obtained with the first-year \emph{WMAP} data, but we do not expect the newer data \cite{spergel06} to improve them substantially given the many uncertainties on the various reionization processes affecting the total optical depth.)
   
However, the CMB is only affected if a substantial fraction of the CMB photons interact with the IGM -- or in other words if the optical depth to electron scattering is substantially altered by the decay products \cite{chen04-decay, pierpaoli04, mapelli06}.  Compared to the IGM temperatures of $\sim 10$--$10^3 \kel$ at $z \sim 30$--$300$, ionization requires a substantial energy input, and decays with long timescales (or low energy injection rates) cannot be ruled out with the CMB.  Nevertheless, because the IGM temperature is so small, these models can still significantly affect the thermal history of the IGM \cite {chen04-decay, mapelli06} and hence the history of star formation. A number of recent particle physics models motivate such scenarios (see the discussion below in \S \ref{particles}).

Dark matter annihilation has also been considered recently, primarily as an explanation for gamma-ray sources in the local Universe (e.g., \cite{profumo06}).  But any such scenario will also predict an annihilation background from encounters between IGM particles; because the physical density increases like $(1+z)^3$, such events could be relatively common in the early Universe and so affect the CMB \cite{padmanabhan05, mapelli06}.  Of course, these scenarios will continue to affect the IGM throughout the dark ages as well.

The 21 cm background can offer  powerful constraints on dark matter decay and annihilation (or indeed any exotic process that injects energy into the IGM during the dark ages).  The hyperfine level populations of the ground state of neutral hydrogen are determined through competition between absorption of (and emission stimulated by) CMB photons (and possibly UV photons) and collisions.  Since the IGM is dense at high-redshifts, collisions dominate, driving the level populations into equilibrium with the kinetic temperature of the gas.  Because the latter is colder than the CMB, the IGM is a net absorber of CMB photons at $z \sim 30$--$300$.  Thus, the 21 cm transition can be used to map fluctuations in the IGM during the dark ages, offering a pristine probe of cosmology \cite{loeb04}.  Previous work has emphasized the possibility of constraining the matter power spectrum with this tool \cite{loeb04, bharadwaj04-vel, barkana05-infall, naoz05}.  Here we point out that the fundamental properties of the fluctuations -- on all scales -- depend sensitively on the thermal history (see also Ref.~\onlinecite{shchekinov06}).  Thus the 21 cm background can be used to constrain  dark matter decay and annihilation.  It is much more sensitive than the CMB to dark matter decay because (i) it depends on the thermal history, not just the ionized fraction, and (ii) it is directly sensitive to the late time behavior (when most of the energy is injected).  It is less useful for constraining annihilation scenarios because in that case a large fraction of the energy is injected at early times.

The excess heating and ionization induced by dark matter decay and annihilation can also affect the chemistry of the IGM.  In the simple chemical environment of the primordial IGM, their most important effect will be on the H$_2$ abundance, which is significant because that molecule is an important coolant for low-temperature gas \cite{saslaw67, galli98} and is thought to play a key role in the formation of the first stars \cite{haiman96, tegmark97, abel02, bromm02}.  Unfortunately, the implications -- and even magnitude -- of any possible boost to early structure formation are controversial \cite{padmanabhan05,biermann06,ripamonti06}.  Here, we provide a more detailed look at ${\rm H_{2}}$ formation and incorporate some hitherto neglected effects in the calculation. 

The rest of this paper is organized as follows.  We briefly review the physics of the 21 cm background in \S \ref{21cm}.  We describe our simple model for dark matter decay and annihilation, and review some particle physics motivation, in \S \ref{dm}.  We then present our results for the 21 cm background in \S \ref{results} and for the H$_2$ abundance in \S \ref{htwo}.  Finally, we discuss their implications in \S \ref{disc}.

In our numerical calculations, we assume a cosmology with $\Omega_m=0.26$, $\Omega_\Lambda=0.74$, $\Omega_b=0.044$, $H=100 h \hunits$ (with $h=0.74$), $n=0.95$, and $\sigma_8=0.8$, consistent with the most recent measurements \citep{spergel06}, although we have increased $\sigma_8$ from the best-fit \emph{WMAP} value to improve agreement with weak lensing.  We quote all distances in comoving units, unless otherwise specified.

\section{\label{21cm}The 21 cm Background}

We review the relevant characteristics of the 21 cm transition here; we refer the interested reader to Ref. \onlinecite{furl06-review} (and references therein) for a more comprehensive discussion.  The 21 cm brightness temperature (relative to the CMB) of a patch of the IGM is
\begin{eqnarray}
\dtb & = & 27 \, \xhi \, (1 + \delta) \, \left( \frac{\Omega_b h^2}{0.023} \right) \left( \frac{0.15}{\Omega_m h^2} \, \frac{1+z}{10} \right)^{1/2} \nonumber \\ 
& & \times \left( \frac{T_S - T_\gamma}{T_S} \right) \, \left[ \frac{H(z)/(1+z)}{\deriv v_\parallel/\deriv r_\parallel} \right] \mkel,
\label{eq:dtb}
\end{eqnarray}
where $\delta$ is the fractional overdensity, $\bxhi = 1 - \xion$ is the neutral fraction, $x_i$ is the ionized fraction, $T_S$ is the spin temperature, $T_\gamma$ is the CMB temperature, and $\deriv v_\parallel/\deriv r_\parallel$ is the gradient of the proper velocity along the line of sight.  When $T_S < T_\gamma$, the IGM appears in absorption.  The last factor accounts for redshift-space distortions  \cite{bharadwaj04-vel,barkana05-vel}.  

Before the first luminous sources  turn on, the spin temperature $T_S$ is determined by competition between scattering of CMB photons, collisions \cite{purcell56}, and scattering of \lya photons \cite{wouthuysen52, field58, hirata05}.  In equilibrium (which is achieved rapidly),
\begin{equation}
T_S^{-1} = \frac{T_\gamma^{-1} + x_c T_K^{-1} + x_\alpha T_c^{-1}}{1 + x_{c} + x_\alpha}.
\label{eq:tsdefn}
\end{equation}
Here $x_c$ is the total collisional coupling coefficient, including both H--H interactions \citep{allison69, zygelman05} and H--e$^-$ collisions \citep{smith66, furl06-elec}.  We will denote the separate coefficients via $x_c^{\rm HH}$ and $x_c^{\rm eH}$, respectively.  The last term describes Wouthuysen-Field coupling:  $x_\alpha$ is the coupling coefficient and $T_c$ is the effective color temperature of the radiation field \cite{chen04, hirata05}.  Typically $T_c \approx T_K$ in the IGM, and 
\begin{equation}
x_\alpha = 1.81 \times 10^{11} (1+z)^{-1} S_\alpha J_\alpha,
\label{eq:xalpha}
\end{equation}
where $S_\alpha$ is a factor of order unity describing the detailed scattering process \cite{chen04, hirata05, chuzhoy06, furl06-lyheat} and $J_\alpha$ is the radiation background at the \lya frequency, in units of photons cm$^{-2}$ s$^{-1}$ Hz$^{-1}$.

From equation (\ref{eq:dtb}) it is obvious that fluctuations in the density, temperature, ionized fraction, radiation background, and velocity all source fluctuations in the brightness temperature.  Because, to linear order in $k$-space, velocity perturbations are simply proportional to density perturbations, we can write the Fourier transform of the fractional 21 cm brightness temperature perturbation as
\begin{equation}
\delta_{21}(\bk) = (\beta + \mu^2) \delta + \beta_H \delta_H + \beta_\alpha \delta_\alpha + \beta_T \delta_T,
\label{eq:d21}
\end{equation}
where $\mu$ is the cosine of the angle between the line of sight and the wavevector $\bk$ and each $\delta_i$ describes the fractional variation in a particular quantity: $\delta_\alpha$ for the \lya coupling coefficient $x_\alpha$, $\delta_H$ for the neutral fraction, and $\delta_T$ for $T_K$.  The expansion coefficients $\beta_i$ are
\begin{eqnarray}
\beta & = & 1 + \frac{x_c}{x_{\rm tot}(1+x_{\rm tot})},
\label{eq:beta} \\
\beta_H & = & 1 + \frac{x_c^{\rm HH} - x_c^{\rm eH}}{x_{\rm tot} (1 + x_{\rm tot})}
\label{eq:betax} \\
\beta_\alpha & = & \frac{x_\alpha}{x_{\rm tot}(1+x_{\rm tot})},
\label{eq:beta-alpha} \\
\beta_T & = & \frac{T_\gamma}{T_K - T_\gamma}  + \frac{x_c}{x_{\rm tot}(1+x_{\rm tot})} \frac{\deriv \ln x_{c}}{\deriv \ln T_K},
\label{eq:betaT} 
\end{eqnarray}
where $x_{\rm tot} \equiv x_c + x_\alpha$.

\section{\label{dm}Dark Matter and the IGM}

The implications of dark matter decay (and/or annihilation) for the thermal and ionization histories of the IGM have been considered by a number of authors \cite{chen04-decay, hansen04, kasuya04, avelino04, pierpaoli04, padmanabhan05, Bean03}.  Here we first review some particle physics scenarios that can affect the dark ages, and then we show how to compute the resulting IGM histories.

\subsection{\label{particles}Some Example Particle Physics Scenarios}

Dark matter particles can ionize and heat the IGM through either decay or annihilation.

For decay, short lifetimes ($t_X \le 3 \times 10^{18} \sec$, with $X$ denoting the dark matter particle) are disfavored, as  they cause the CMB power spectra to disagree with existing data \cite{chen04-decay, pierpaoli04}.  On the other hand, particles with exceptionally long lifetimes, such as gravitinos with mass $m_X c^2 \simeq 10$--$100 \, {\rm MeV}$ (and hence $t_X \simeq 10^{31} \sec$ \cite{HW04}) only affect the IGM temperature at low redshift ($z \le 2$) \cite{mapelli06}.  We wish to consider  particles with lifetimes in the range  $t_X \sim 10^{24}$--$10^{27} \sec$. 

One such example is an axino with mass $m_X c^2 \sim 1$--$100~ {\rm MeV}$; positrons produced as their decay products could explain the 511 keV emission excess from the Galactic center \cite{HW04}.  Axinos in this mass range have $t_X \simeq 3 \times 10^{24}$--$10^{26} ({\rm MeV}/m_X c^2) \sec$.  The ionized fraction produced by such particles is too low to be observable with the CMB \cite{mapelli06}, but the IGM temperature departs from the standard case by $z \simeq 100$ and at $z \simeq 20$ is $\sim 20$ times  higher, if energy transfer is perfectly efficient \cite{mapelli06}.  However, in these particular models a detailed accounting of the interactions between decay products and the IGM substantially decreases their effect and pushes it to lower redshifts \cite{ripamonti06}.

Another possible candidate  is a sterile neutrino with $m_X c^2 \sim 2$--$4 \keV$.  Again, even for the highest mass in this range, the ionized fraction produced by these particles is small ($x_i \le 0.1$ at $z=0$), but the present-day IGM temperature increases by several orders of magnitude (although again the energy transfer is likely not perfectly efficient; \cite{mapelli06, ripamonti06}).  Although such a neutrino is unlikely to be the sole dark matter component \cite{Seljak}, it may still constitute a  non-negligible fraction of the dark matter.  Such neutrinos have decay times $t_X \simeq 3 \times 10^{27} \sec$.

Decays of super-heavy  dark matter  particles ($m_X c^2 \ge  10^{12} ~{\rm GeV}$) have been considered as possible sources of reionization \cite{Doro02,Doro03} and also as a production mechanism for ultra-high energy cosmic rays \cite{BKV97,BS98}.  With such energetic particles, the precise ionization and heating strongly depend on the type of reactions assumed.

Two kinds of annihilating dark matter have been investigated in connection with non-standard reionization:  neutralino annihilation (e.g. \cite{GK00,BHS04}), involving particles with $m_X c^2 \ge 30 ~{\rm GeV}$, and light dark matter annihilation ($m_X c^2 \sim 1$--$100 ~{\rm MeV}$ \cite{Boehm04}).  While the expected ionization fraction is small, it extends to high redshifts, greatly affecting the CMB power spectrum.   These models can already be constrained with current  CMB data and   will be better constrained by future CMB polarization measurements \cite{padmanabhan05, mapelli06}.  In the most reasonable models, the IGM temperature can be several times larger than normally expected at $z \sim 20$--$30$.

The effects of light dark matter annihilation on CMB power spectra have also been considered.  CMB data still allow models with small ionization fractions but with IGM temperatures a few times larger than in the standard scenario (e.g., $\sim 2$ times larger at $z=60$; \cite{zhang06, ripamonti06}).

\subsection{Energy Deposition}

Rather than examine each of these models individually, we will follow the method of Ref.~\onlinecite{chen04-decay}, who presented a simple but general model for the effects of dark matter decay.  We first suppose that the dark matter particle has a decay rate $\Gamma_X=t_X^{-1}$, so that the physical number density of particles is
\begin{equation}
n_X(z) = n_X^0 \, (1+z)^3 \, e^{-\Gamma_X t},
\label{eq:ndm}
\end{equation}
where $n_X^0$ is the comoving number density.  In our calculations we will assume $\Gamma_X \ll H_0$, so that the exponential factor can be ignored.  While not strictly necessary, this is probably the most interesting case (because otherwise the decaying particle could not serve as the dark matter we see around us today) and serves to illustrate the basic results.  The total comoving emissivity is then
\begin{equation}
Q^{\rm thick} = f \Gamma_X \, n_X \, m_X c^2,
\label{eq:qthick}
\end{equation}
where $f$ is the fraction of decay energy that is potentially available to the IGM (e.g., excluding particles that free stream without interacting, such as neutrinos); the meaning of the superscript will become apparent momentarily.  It is convenient to define $\xi_X \equiv \Gamma_X \Omega_X/\Omega_b \approx 5.9 \Gamma_X$ to be the decay rate normalized for notational convenience to the baryon fraction (i.e., the equivalent decay rate if it were the baryons that decayed).  Then $Q^{\rm thick} = f \xi_X n_b^0 m_p c^2$, where $n_b^0$ is the comoving number density of baryons.

For concreteness, we will assume that the particle decays to high-energy photons, which then either scatter through the IGM or free stream to lower redshifts.  (Other decay products are of course possible and will have similar qualitative effects; the important parameter is the total energy injection rate.)  High-energy photons can interact with their surroundings through photoionization, Compton scattering, pair production (via scattering off neutral atoms, free electrons, CMB photons, or ions), and scattering with CMB photons \cite{zdziarski89}.  At the high redshifts of interest here, photoionization and scattering off of free electrons and atoms dominates energy loss at low photon energies ($E \la 3 \keV$); in this regime, the IGM is opaque.  At high energies ($E \ga 10$--$10^3$ GeV, depending on redshift), pair production off CMB photons dominates, and the total optical depth is also large.  The resulting pairs rapidly deposit their energy in the IGM.  However, between these thresholds lies a ``transparency window" of relatively low optical depth; photons in this regime experience optical depths (over a Hubble length) $\tau \sim 10^{-2}$--$1$ at $z=100$ \cite{chen04-decay}.  Thus they lose only a fraction of their energy to the IGM, with the rest free streaming to the present or being lost to cosmic expansion.

The decay energy can be deposited in heating, collisional excitation, or ionization; while the fractions $\chi_h$, $\chi_e$, and $\chi_i$ in each process actually depend on the initial photon energy and the ionized fraction \cite{shull85}, to the accuracy of our simple models the rules $\chi_i \sim \chi_e \sim (1-\xion)/3$ and $\chi_h \sim (1 + 2 x_e)/3$ suffice \cite{chen04-decay}. 

When the decay products experience a large optical depth, the energy deposition rate is simply $Q^{\rm thick}$ (hence the superscript).  However, if the energy is injected in the transparency window, we have instead
\begin{equation}
Q^{\rm thin} = Q^{\rm thick} [1-e^{-\tau(z)}] \approx Q^{\rm thick} \tau(z),
\label{eq:qthin}
\end{equation}
where $\tau(z) = c n_b(z) \sigma_{\rm eff}/H(z)$ is the effective optical depth for interactions over a Hubble length.  Here $\sigma_{\rm eff}$ is the energy-averaged cross-section for interactions.  Rather than attempt to compute the relevant processes in detail, we will write
\begin{equation}
\tau(z) \equiv \tau_{\rm 100} \left( \frac{1+z}{100} \right)^{3/2},
\label{eq:taueff}
\end{equation}
where $\tau_{\rm 100}$ is the optical depth at $1+z=100$ and the exponent assumes a constant $\sigma_{\rm eff}$ (so that $\tau \propto n_b/H$).  More detailed calculations of $\tau(z)$ can take into account the specific interactions with the IGM for a given particle physics model (e.g., \cite{ripamonti06}), but we forego these here to keep our calculations generic.  Thus decay into the transparency window can be described in the same way as instantaneous energy injection provided we make the replacement $\xi_X \rightarrow \xi_X [1-{\rm exp}(-\tau)]$.  In this case, most of the energy would free-stream to the present day and contribute to the X-ray and $\gamma$-ray backgrounds.  If the decay photons are monoenergetic, the observed background provides fairly powerful constraints: $\xi_X \la 10^{-25} \secinv$ for $m_X c^2 \la 10$ MeV or $\xi_X \la 10^{-30} \secinv$ for $m_X c^2 =100$ GeV \cite{chen04-decay}.  However, existing constraints allow $\xi_X$ to be much larger if the photons have a wide range of energies.

We will also consider the case of dark matter annihilation.  Neglecting the evolving clumpiness of the dark matter field, the comoving energy deposition rate is
\begin{equation}
Q^{\rm ann}= 2 f \, m_X c^2 \, (n_X^0)^2 \VEV{\sigma v} (1+z)^3,
\label{eq:qann}
\end{equation}
where the factor of two occurs because two particles annihilate in each event and $\VEV{\sigma v}$ is the velocity-averaged annihilation cross section.  In this case, we can define an effective baryon-normalized ``lifetime" via
\begin{equation}
\xi_X^{\rm ann} \equiv \frac{\Omega_X \rho_c^0}{m_X} \VEV{\sigma v} (1+z)^3 \left( \frac{\Omega_X}{\Omega_b} \right).
\label{eq:xiann}
\end{equation}
Like decay into the transparency window, annihilation can be thought of as a redshift-dependent decay rate.  Obviously, annihilation injects relatively more energy at high redshifts.  For particle decay, the free parameter is $m_X \Gamma_X$, which is easy to interpret.  For annihilation, it is the less intuitive quantity $\VEV{\sigma v}/m_X$.  Reasonable values for neutralino annihilation are $\VEV{\sigma v} \approx 2 \times 10^{-26} \recunits$ and $m_X c^2 = 100 \GeV$ (e.g., \cite{padmanabhan05}).  For illustrative purposes, we will consider a range of cross sections up to $\VEV{\sigma v} \approx 2.5 \times 10^{-25} \recunits$ (with $f=1$) -- somewhat above the upper limit provided by existing CMB data \cite{mapelli06}. (Note that our value of $\VEV{\sigma v}$ is four times smaller than that used by Ref. \onlinecite{mapelli06}, because they assumed $f=0.25$.)

\subsection{The Mean Evolution of the IGM}

We now wish to describe how these processes affect the IGM; our approach in this section is similar to Ref.~\onlinecite{shchekinov06}.  First consider the effects on the globally-averaged temperature and ionized fraction.  Without dark matter, the mean IGM temperature $\bar{T}_K$ is determined by the competition between adiabatic cooling and Compton scattering of CMB photons off of the residual free electrons.  We obtain \cite{seager99, chen04-decay, pierpaoli04}
\begin{eqnarray}
\frac{\deriv \bar{T}_K}{\deriv t} & = & - 2 \bar{T}_K H(z) + \frac{x_i(z)}{\eta_1 t_\gamma} \, (T_\gamma - \bar{T}_K) \nonumber \\ & & - \frac{2}{3} \, \frac{\eta_2 m_p c^2}{\eta_1 k_B} \, \xi_X \chi_h,
\label{eq:tk}
\end{eqnarray}
where $t_\gamma=(3 m_e c)/(8 \sigma_T u_{\rm CMB})$, $u_{\rm CMB}$ is the CMB energy density at redshift $z$, $\eta_1 = (1 + f_{\rm He} + \bxion)$, $\eta_2 = (1 + 4 f_{\rm He})$, and $f_{\rm He}$ is the helium fraction by number (we assume helium remains neutral for simplicity)

The mean ionized fraction evolves according to
\begin{equation}
\frac{\deriv \bxion}{\deriv t} = -\alpha \bxion^2 \bar{n}_H + \eta_2 \left( \frac{m_p c^2}{E_{\rm ion}} \right) \xi_X \chi_i,
\label{eq:xionevol}
\end{equation}
where the first term describes recombinations with an effective coefficient $\alpha$ (taken from Ref. \onlinecite{seager99}) and $\bar{n}_H$ is the total mean density of ionized and neutral hydrogen nuclei (note that gas clumping is negligible at the redshifts of interest here).  The second term results from dark matter decay; $E_{\rm ion}=13.6 \eV$ is the hydrogen ionization threshold.  Most ionizations are caused by collisions with hot secondary photo-electrons rather than by direct photoionization, with a total number of ionizations $\sim E_\gamma/(37 \eV)$ for a photon of energy $E_\gamma$ at  $\bxion \ll1$ \cite{shull85} -- which agrees well with our assumption that $\chi_{i}\sim (1-x_{i})/3$.  Note that, by increasing $\bxion$, dark matter decay also makes Compton scattering more efficient in equation~(\ref{eq:tk}), which provides a second channel for heat input.  This modification to the reionization history affects the CMB temperature and polarization power spectra through Thomson scattering during the dark ages.  The \emph{WMAP} experiment requires $\xi_X \la 10^{-24} \secinv$ \cite{chen04-decay, pierpaoli04} for energy deposition in the optically thick regime.

A more subtle effect of the energy injection is to set up a global Ly$\alpha$ background, which can couple the spin and kinetic temperatures through the Wouthuysen-Field effect.  This occurs through collisional excitations induced by fast photoelectrons (see the analogous discussions in the context of X-ray heating in \cite{chen06, chuzhoy06-first, pritchard06}).  Each hydrogen excitation will produce a cascade of line photons; the Lyman-series photon produced in the cascade will then scatter through the (extremely optically thick) IGM.  Higher Ly$n$ photons will eventually cascade further, with $\sim 1/3$ of them being ``recycled" into \lya photons \cite{hirata05, pritchard05}.  If a fraction $\chi_\alpha$ of the input energy eventually goes into \lya photons, the background is
\begin{equation}
J_\alpha \approx \frac{c}{4\pi} \, \frac{m_p c^2}{h \nu_\alpha^2} \, \frac{\chi_\alpha \xi_X n_b}{H(z)},
\label{eq:jalpha}
\end{equation}
where $\nu_\alpha$ is the \lya line frequency and $n_b$ is the number density of baryons.  Interestingly, this can induce relatively large coupling:
\begin{equation}
x_\alpha \sim 4.5 \,  \xi_{-24} \, S_\alpha \, \left( \frac{\chi_\alpha}{1/6} \right) \, \left( \frac{1+z}{100} \right)^{1/2},
\label{eq:xalpha-dm}
\end{equation}
where $\xi_{-24} \equiv \xi_X/(10^{-24} \sec)$.  As we shall see, this becomes potentially important at lower redshifts, when collisional coupling between $T_{\rm S}$ and $T_{\rm K}$ weakens.  Estimating the precise coupling requires a detailed examination of the collisional excitation processes, which we will forego here given our generic and crude approach.  We will simply assume that one-half of the total excitation energy goes into \lya photons, i.e. $\chi_\alpha = \chi_e/2$ (see \cite{pritchard06} for a more comprehensive discussion).  This \lya background also affects the IGM temperature, but the rates are completely negligible in the models we consider \cite{chen04, chuzhoy06-heat, furl06-lyheat}.

\subsection{Fluctuations in the IGM}

Equations~(\ref{eq:tk}), (\ref{eq:xionevol}), and (\ref{eq:jalpha}) suffice to describe the evolution of the global 21 cm background.  But a more interesting observable is how this signal fluctuates, which is easier to measure because the signal monopole is severely contaminated by foregrounds
(see \cite{furl06-review} and references therein).  Equation~(\ref{eq:d21}) shows that these fluctuations have a variety of sources.  How do variations in the energy deposition rate source 21 cm fluctuations?  We can imagine some simple cases for perturbations to $Q$.  The first is uniform deposition.  This would be appropriate if, for example, the decays produce particles with long mean-free paths that then deposit energy after interacting with a uniform background (such as the CMB).  This is probably not realistic, because in most cases the majority of energy is ultimately transferred to the IGM through relatively low-energy photons or particles with short mean free paths.  In such a scenario, the local energy injection rate will be proportional to the density (because the energy deposition rate \emph{per particle} is constant), even if the intermediate decay products have reasonably long mean free paths.  To distinguish these scenarios, we define a variable $\theta_u$ such that the energy deposition rate is proportional to $(1 + \theta_u \delta)$ (i.e., $\theta_u=0$ for uniformly deposited energy).  For annihilating particles, $Q \propto n_X^2 \propto (1+\delta)^2$.  If the energy is deposited immediately, we would have $\theta_u=2$ (assuming that $\delta \ll 1$ and Taylor expanding); in the perhaps more plausible case that the intermediate products have long mean free paths, $\theta_u=1$ would be more appropriate.

The fractional perturbation to the gas temperature, $\delta_T$, obeys (cf., \cite{bharadwaj04-vel, naoz05}) 
\begin{eqnarray}
\frac{\deriv \delta_T}{\deriv t} & = & \frac{2}{3} \, \frac{\deriv \delta}{\deriv t} - \frac{\bxion}{\eta_1 t_\gamma} \, \frac{T_\gamma}{\bar{T}_K} \, \delta_T + \frac{\bxion \delta_i}{t_\gamma} \frac{T_\gamma - \bar{T}_K}{\bar{T}_K} \nonumber \\
& & - \frac{2}{3} \, \frac{\eta_2 m_p c^2}{\eta_1 k_B \bar{T}_K} \, \xi_X \chi_h [(1-\theta_u) \delta +\delta_T],
\label{eq:deltaT}
\end{eqnarray}
where $\delta_i = - \delta_H (1 - \bxion)/\bxion$ is the fractional perturbation to the ionized fraction.  
The first term describes adiabatic compression or expansion, the second and third describe fluctuations in the rate at which energy is transferred through Compton scattering, and the last term describes how dark matter decay affects the fluctuations.  (Note that we have ignored terms of order $\bxion \delta$, because the ionized fraction remains small in all of the models we consider.  We have also ignored variations in $\chi_h$ and $\chi_i$ with the ionized fraction.)  The corresponding equation for $\delta_i$ is
\begin{eqnarray}
\frac{\deriv \delta_i}{\deriv t} & = & - \alpha \bxion \bar{n}_H ( \delta_i + \delta + \alpha' \delta_T ) \nonumber \\ 
& & - \eta_2 \left( \frac{m_p c^2}{E_{\rm ion}} \right) \frac{\xi_X \chi_i}{\bxion} [\delta_i + \delta(1 - \theta_u)],
\label{eq:deltai}
\end{eqnarray}
where the first term describes fluctuations in the recombination rate (which is a function of temperature; we have let $\alpha' \equiv \deriv \ln \alpha/\deriv \ln T_K$) and the second accounts for dark matter decays.  Note that, in the absence of extra energy injection, $\delta_i \approx 0$ is a reasonable approximation at $z \la 150$ (cf. \cite{barkana05-infall}), because recombinations are slow in the largely neutral gas, but it must be included in the more general case.

The appearance of $\delta$, $\delta_T$, and $\delta_i$ in each of these equations implies that the different density modes source independent temperature and ionization modes.  This introduces a non-trivial scale dependence into the calculation \cite{barkana05-infall, naoz05}.  But for a simple estimate it suffices to follow only the growing density mode, $\delta \propto a$, which dominates on the large scales that may ultimately be observable \cite{bharadwaj04-vel}.  Writing $\delta_T \equiv g_T(z) \delta$ and $\delta_i \equiv g_i(z) \delta$, we have 
\begin{eqnarray}
\frac{\deriv g_T}{\deriv z} & \approx & \frac{g_T - 2/3}{1+z} + \frac{\bxion}{\eta_1 t_\gamma} \, \frac{g_T T_\gamma - g_i (T_\gamma -\bar{T}_K)}{\bar{T}_K (1+z) H(z)} \nonumber \\
& & + \frac{2}{3} \, \frac{\eta_2 m_p c^2}{\eta_1 k_B \bar{T}_K}  \, \xi_X \chi_h \frac{(1 - \theta_u) + g_T}{(1+z) H(z)},
\label{eq:gtz} \\
\frac{\deriv g_i}{\deriv z} & \approx & \frac{g_i}{1+z} + \frac{\alpha \bxion \bar{n}_H (1 + g_i + \alpha' g_T)}{(1+z) H(z)} \nonumber \\
& & + \eta_2 \left( \frac{m_p c^2}{E_{\rm ion}} \right) \frac{\xi_X \chi_i}{\bxion} \frac{(1-\theta_u)+ g_i}{(1+z)H(z)}.
\label{eq:giz}
\end{eqnarray}
These equations make the effects of dark matter perturbations obvious.  First consider $g_T$.  Adiabatic expansion and compression tend to drive $\delta_T \rightarrow 2\delta/3$.  But when Compton cooling is efficient (and if $\delta_i=0$), it inputs a constant amount of energy per particle, driving the gas toward isothermality ($g_T \rightarrow 0$).  Decay or annihilation energy that is injected in the same way (proportional to the local gas density) also tends to drive the gas toward isothermality.  On the other hand, if the energy is injected uniformly, it drives $g_T \rightarrow -1$ because it must be shared between more particles in denser regions.

In equation~(\ref{eq:giz}), the first term holds $\delta_i$ constant in the absence of recombinations.  The second term shows that recombinations tend to drive $g_i \rightarrow -1$, because denser gas recombines more quickly.  The dark matter decay term behaves similarly to its analog in equation~(\ref{eq:gtz}):  injection at a constant rate per particle damps out perturbations in $\xion$, while uniform injection preferentially decreases the ionized fraction in dense regions.

Finally, we must consider fluctuations in the \lya background.  If the energy is injected uniformly, $J_\alpha$ is constant and $\delta_\alpha=0$.  If $\theta_u=1$, $\delta_\alpha=\delta$.  Note the contrast with $\delta_T$ and $\delta_i$: the coupling efficiency depends on the total \lya background, \emph{not} the background per particle, because each \lya photon scatters many times.

\begin{figure}
\resizebox{8cm}{!}{\includegraphics{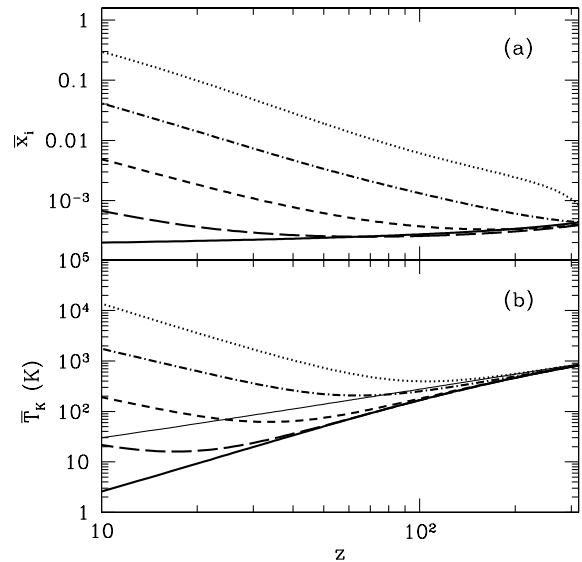}}
\caption{\label{fig:therm-thick} IGM histories for long-lived dark matter.  In each panel, the curves take $\xi_{-24}=1,\,0.1,\,10^{-2},\,10^{-3}$, and $0$, from top to bottom.  \emph{(a)}: Ionization histories.  \emph{(b)}: Thermal histories.  Here the thin solid curve shows $T_\gamma$. }
\end{figure}

Using the baryonic power spectrum $P_{\delta \delta}(k)$ and integrating equations~(\ref{eq:gtz}) and (\ref{eq:giz}), we can estimate the fluctuation amplitude for $\delta_T$, $\delta_i$, and $\delta_\alpha$ as a function of scale at any redshift.  The power spectrum of the 21 cm fluctuations is \cite{furl06-review}
\begin{equation}
P_{21}(k,\mu) = \bdtb^2 \, (\beta' + \mu^2) P_{\delta \delta}(k),
\label{eq:pk}
\end{equation}
where 
\begin{equation}
\beta' = \beta + \beta_T g_T - \beta_H \bxion g_i/(1 - \bxion) + \theta_u \beta_\alpha.
\label{eq:betaprime}
\end{equation}
For simplicity, we will average over the $\mu$ dependence when presenting our results.  Note that, because the 21 cm power spectrum is simply proportional to the matter power spectrum on large scales, we will only present results at a single wavenumber $k=0.04 \Mpcinv$.  The shape is straightforward to compute \cite{loeb04, barkana05-infall}, but we are concerned with the differences induced by dark matter decay and annihilation.  Because we have used only the growing mode, these differences are scale-independent.  We have chosen $k=0.04 \Mpcinv$ because it is in the range of scales most accessible to observations (see \S \ref{disc}).

\section{\label{results}Results:  The 21 cm Background}

\subsection{Dark Matter Decay: Optically Thick Regime \label{thick}}

We have integrated the temperature and ionization equations with initial values taken from RECFAST \cite{seager99} at $z=350$ (dark matter decay at higher redshifts does not significantly affect the history).  Figure~\ref{fig:therm-thick} shows the resulting thermal and ionization histories for $\xi_{-24} = 1,\,0.1,\,10^{-2},\,10^{-3}$ and $0$ from top to bottom.  These correspond to lifetimes $t_X \sim 6 \times (10^{24}$--$10^{27}) \sec$, assuming that the decaying particle makes up a large fraction of the dark matter.  The uppermost curve would significantly affect the CMB and can be ruled out by WMAP \cite{chen04-decay, pierpaoli04}, but the others have almost no effect on it (the total optical depths to electron scattering from $z>10$ are $\tau_{\rm es}=0.083,\,0.016,\,0.0060,\,0.0048$ and $0.0046$ in these models).

The most important point of Figure~\ref{fig:therm-thick} is that, even if $\bxion$ remains small, $\bar{T}_K$ can still increase significantly.  This is of course simply because the IGM is so cold at these redshifts:  if roughly equal amounts of energy go toward ionization and heating, we would expect $\bar{T}_K \sim 10^3 (\bxion/0.1) \kel$, far above the usual temperature.  For the larger values of $\xi_X$, the increased efficiency of Compton scattering in the more highly-ionized Universe exaggerates the effectiveness of heating.  This aspect is invisible so far as the CMB is concerned, but the 21 cm background measures it directly. 

\begin{figure}
\resizebox{8cm}{!}{\includegraphics{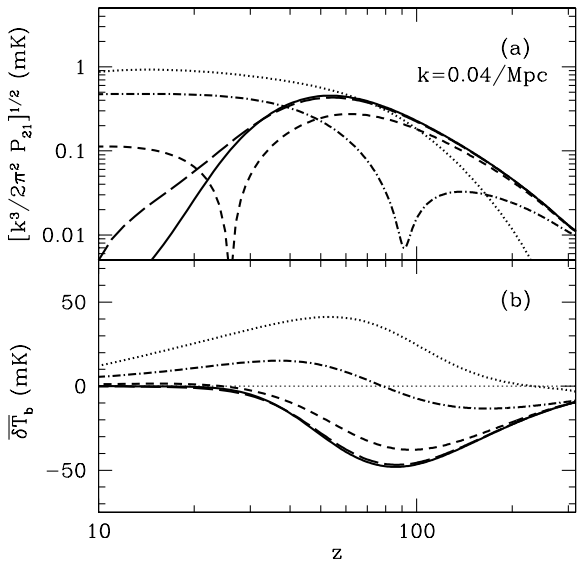}}
\caption{\label{fig:signal-thick} 21 cm signals for long-lived dark matter.  Curves take the same parameters as in Fig.~\ref{fig:therm-thick}.  \emph{(a)}: Fluctuation amplitude at $k=0.04 \Mpcinv$ (note that this scale is arbitrary). \emph{(b)}: Mean (sky-averaged) signal. The thin horizontal dotted line shows $\bdtb=0$.}
\end{figure}

Figure~\ref{fig:signal-thick} shows the resulting 21 cm signals.  Panel \emph{(a)} shows the angle-averaged fluctuation amplitude at $k=0.04 \Mpcinv$ in the different decay scenarios of Figure~\ref{fig:therm-thick}.  In this (and in all figures unless otherwise specified), we assume $\theta_u=1$ (i.e., the energy deposition rate is proportional to the local density).  Panel \emph{(b)} shows the sky-averaged 21 cm brightness temperature as a function of redshift.

First consider the solid curves, which present the standard calculation without dark matter decay (e.g., \cite{loeb04, bharadwaj04-vel}).  In this case, Compton scattering becomes inefficient at $z \sim 200$, and the gas begins to cool below $T_\gamma$.  At $z \ga 80$, the density is large enough for collisions to drive $T_S \rightarrow T_K$, so the IGM becomes visible in absorption.  Below this redshift, $\dtb$ returns to zero because collisional coupling becomes inefficient as the density decreases.  The fluctuation amplitude behaves similarly, except with a peak at $z \sim 50$, because it is weighted by the growing density mode, $\delta \propto (1+z)^{-1}$ \cite{loeb04, bharadwaj04-vel}.

The effects of dark matter decay on $\bdtb$ are straightforward and compare well to Ref.~\onlinecite{shchekinov06}:  by continually heating the IGM, it decreases the intensity of absorption or even turns it into emission.  In the strongest decay models, the extra heating is sufficient to render the IGM visible at $z \sim 10$ even without luminous sources:  this is because the collisional coupling rate $x_c$ is a sensitive function of $\bar{T}_K$.  Nevertheless, the peak signal still occurs at much higher redshifts. Interestingly, the fluctuations evolve in a non-trivial way.  At the highest redshifts, $P_{21}$ is always small. But it is weakest when dark matter decay is strongest, because the increased ionized fraction helps keep $\bar{T}_K \approx T_\gamma$.  In all of these scenarios, the rms amplitude reaches mK levels by $z \sim 50$.  With strong heating, it remains large at lower redshifts even though $\bdtb$  decreases, because (i) the fluctuations continue growing, (ii) the increase in $\bar{T}_K$ makes collisional coupling more efficient \cite{zygelman05}, and (iii) the \lya background continues to contribute.

\begin{figure}
\resizebox{8cm}{!}{\includegraphics{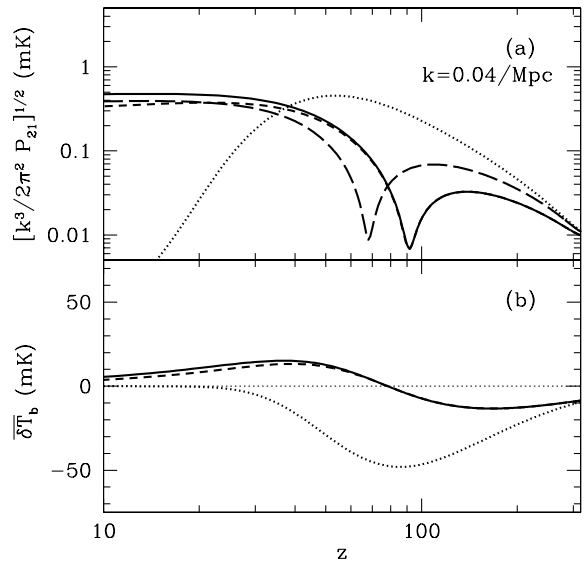}}
\caption{\label{fig:breakdown} As Fig.~\ref{fig:signal-thick}.  The dotted curves take $\xi_{-24}=0$; the rest have $\xi_{-24}=0.1$.  The solid curves show the net signal.  The short-dashed curves set $\chi_\alpha=0$.  The long-dashed curve (shown only in the top panel) assumes $\theta_u=0$.  }
\end{figure}

The cases with moderately strong heating show the most interesting structure, because in such scenarios the 21 cm signal changes from absorption to emission.  Near to (but slightly before) the crossover point (at which $\bar{T}_K=T_\gamma$), $P_{21}$ also goes to zero.  (When $\bdtb=0$, overdense regions have $\bar{T}_K > T_\gamma$ and hence are visible.)  This transition point would be a clear signature of strong heating from some exotic process.  Unfortunately, in many models the clearest differences occur at $z \la 40$, when confusion with (rare) luminous sources may make it difficult to separate the signal.

\begin{figure}
\resizebox{8cm}{!}{\includegraphics{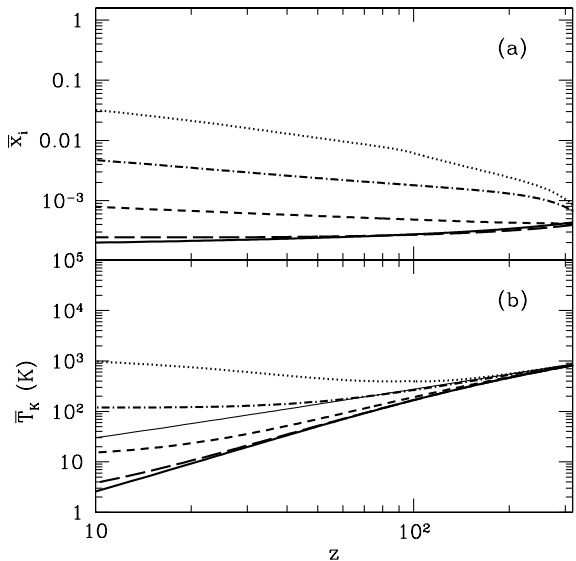}}
\caption{\label{fig:therm-thin} As Fig.~\ref{fig:therm-thick}, except for energy injection in the transparency window.  The bottom solid curve in each panel assumes no extra energy injection.  The other curves take $\xi_{-24}=1$.  The dotted, dot-dashed, short-dashed, and long-dashed curves take $\tau_{100}=1,\,10^{-1},\,10^{-2}$ and $10^{-3}$, respectively.}
\end{figure}

Figure~\ref{fig:breakdown} illustrates how the different processes shape the curves in Figure~\ref{fig:signal-thick} for a fiducial model with $\xi_{-24}=0.1$ (solid curve). Here the dotted curve assumes the standard recombination history ($\xi_{-24}=0$).  The short-dashed curve ignores the \lya photons created through collisional excitation.  This makes no difference at high redshifts, where collisional coupling is already efficient, but decreases the mean signal by $\sim 25\%$ and the rms fluctuations by $\sim 20\%$ at low redshifts.   In this regime the \lya background helps to maintain contact between $T_S$ and $T_K$ (even though $x_\alpha \propto [1+z]^{1/2}$).  The long-dashed curve assumes $\theta_u=0$, so that the energy is deposited uniformly.   (Spatial fluctuations in the dark matter decay rate obviously have no effect on $\bdtb$, so we do not show this curve in the bottom panel.)  This \emph{increases} the fluctuation amplitude at higher redshifts by decreasing the heating rate in dense gas so that it absorbs more strongly.  For the same reason, the ``zero-point" for the fluctuations actually \emph{follows} that of $\bdtb$.  Of course, once the gas appears in emission, uniform energy injection tends to damp out the fluctuations (because dense regions remain colder and hence less luminous).

\begin{figure}
\resizebox{8cm}{!}{\includegraphics{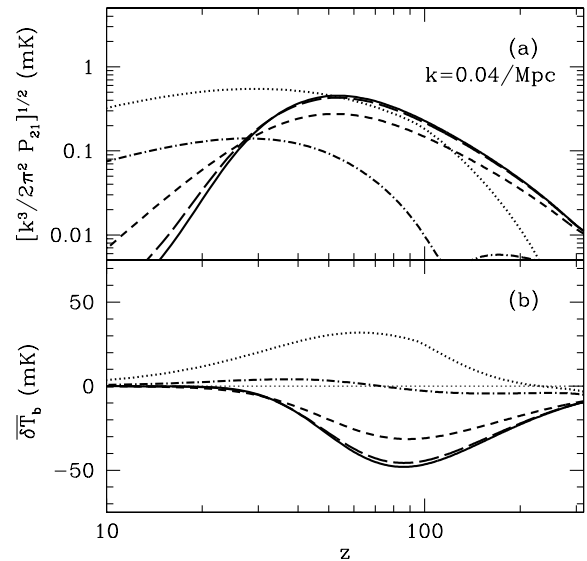}}
\caption{\label{fig:signal-thin} 21 cm signals for long-lived dark matter with energy injection in the transparency window.  Curves take the same parameters as in Fig.~\ref{fig:therm-thin}.  \emph{(a)}: Fluctuation amplitude at $k=0.04 \Mpcinv$ (note that this scale is arbitrary). \emph{(b)}: Mean (sky-averaged) signal. }
\end{figure}

\subsection{Dark Matter Decay:  The Optically Thin Regime \label{thin}}

Figure~\ref{fig:therm-thin} shows the thermal and ionization histories for several scenarios where energy is deposited in the transparency window.  We take $\xi_{-24}=1$ and assume a roughly constant cross section.  The dotted, dot-dashed, short-dashed, and long-dashed curves take $\tau_{100} =1,\,10^{-1},\,10^{-2}$ and $10^{-3}$, respectively; note that at any given redshift $\tau$ and $\xi_X$ are essentially degenerate in our simple model:  the net energy deposition rates at $z=100$ in these models precisely equal the corresponding curves in Figure~\ref{fig:therm-thick}.  It is only the redshift evolution that changes.  We note that $\tau_{\rm es}=0.047,\,0.018,\,0.0062$, and $0.0046$ for these models (including only gas at $z>10$).  Thus the dotted curve substantially affects the CMB (and can already be ruled out), but the others have quite weak effects on it.

For a fixed effective energy deposition rate, the major difference within the transparency window is that a larger fraction of the heating and ionization occurs at relatively high redshifts, with $\bxion$ and $\bar{T}_K$ increasing more slowly at lower redshifts.  As a result the crossover point $\bar{T}_K > T_\gamma$ occurs earlier (if it occurs at all, of course), so the features in the 21 cm signal (shown in Figure~\ref{fig:signal-thin}) occur at significantly lower frequencies.  Nevertheless, they have the same general structure compared to the standard calculation, with a reduced fluctuation amplitude at higher redshifts but stronger fluctuations at lower redshift.  The overall magnitude of the effect is comparable to the optically thick models in Figure~\ref{fig:signal-thick}, although here the observable consequences tend to be stronger at higher redshifts and weaker at lower redshifts.

\subsection{Dark Matter Annihilation \label{ann-res}}

\begin{figure}
\resizebox{8cm}{!}{\includegraphics{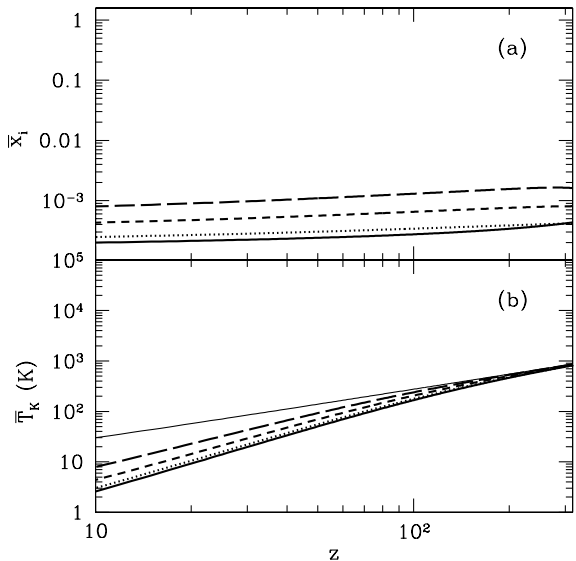}}
\caption{\label{fig:therm-ann} As Fig.~\ref{fig:therm-thick}, except for dark matter annihilation.  The bottom solid curve in each panel assumes no extra energy injection.  The others take $m_X c^2=100 \GeV$ and $\VEV{\sigma v}= 1.25,\, 5.6$, and $25 \times 10^{-26} \recunits$ (dotted, short-dashed, and long-dashed curves, respectively).}
\end{figure}

\begin{figure}
\resizebox{8cm}{!}{\includegraphics{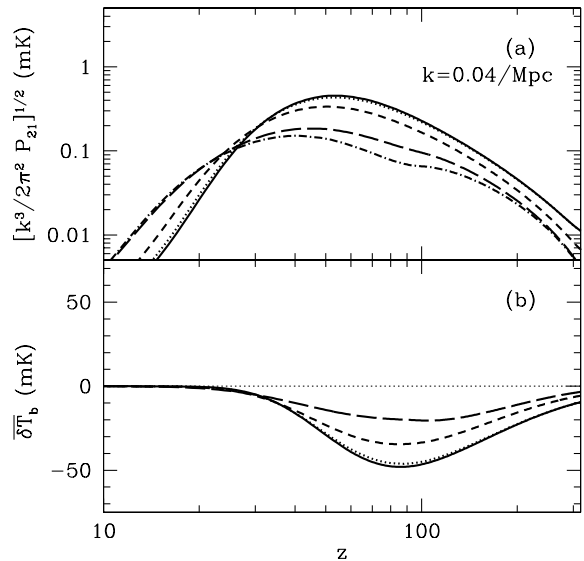}}
\caption{\label{fig:signal-ann} 21 cm signals for annihilating dark matter.  Most curves take the same parameters as in Fig.~\ref{fig:therm-ann}, with $\theta_u=1$.  The exception is the dot-dashed curve in the top panel, which is identical to the long-dashed curve except that it assumes $\theta_u=2$.  \emph{(a)}: Fluctuation amplitude at $k=0.04 \Mpcinv$ (note that this scale is arbitrary). \emph{(b)}: Mean (sky-averaged) signal. }
\end{figure}

Figures~\ref{fig:therm-ann} and \ref{fig:signal-ann} show results for several models of dark matter annihilation, with $m_X c^2 = 100 \GeV$ and $\VEV{\sigma v}= 1.25,\, 5.6$, and $25 \times 10^{-26} \recunits$ (from bottom to top).  The uppermost curve corresponds to the strong annihilation case considered by Ref.~\onlinecite{mapelli06} and is near the upper limit set by \emph{WMAP}.  As expected from $\xi_X^{\rm ann} \propto (1+z)^3$, most of the energy is deposited at high redshifts.  In Figure~\ref{fig:therm-ann}\emph{a}, this manifests itself as an elevated ionization fraction at $z \gg 100$, which decreases slowly toward lower redshift because $\xi_X^{\rm ann}(z)$ becomes almost negligible and recombinations take over.  The effect on $\bar{T}_K$ is much smaller at high redshifts because most of the energy is deposited while Compton scattering is still efficient, which is much stronger than dark matter annihilation.  On the other hand, at lower redshifts the gas has cooled sufficiently that even the tiny energy injection rate from annihilations suffices to increase the temperature by up to a factor of a few.

Figure~\ref{fig:signal-ann} shows that the effects on $\bdtb$ and the 21 cm fluctuations are also relatively modest.  The strong annihilation case decreases both the mean signal and the fluctuation amplitude at high redshifts by a factor of several, because the IGM temperature remains closer to $T_\gamma$.  It modestly increases the fluctuations at lower redshifts because of the increased $\bar{T}_K$; at temperatures $\la 100 \kel$, the collisional coupling efficiency is quite sensitive to temperature \cite{zygelman05}.  Unfortunately, more realistic scenarios with weaker $\VEV{\sigma v}$ have considerably smaller effects that will be difficult to observe given the challenges posed by the observations; they are at best comparable to some of the slower decay models above.  

As with dark matter decay, perturbations in the energy injection rate affect $\delta_T$ and $\delta_i$ (and hence the 21 cm fluctuation signal).  Most of the curves in Figure~\ref{fig:signal-ann}\emph{a} assume $\theta_u=1$, or in other words that the energy injection rate is proportional to the local density.  But of course the annihilation rate is actually proportional to $n_X^2$; if the products have a short mean free path and interact with the local IGM, we would have $\theta_u=2$.  This is shown for $\VEV{\sigma v}=25 \times 10^{-26} \recunits$ by the dot-dashed curve.  Comparison to the long-dashed curve shows that this scenario further suppresses the high-redshift fluctuations, because it injects even more energy into the dense spots.  Their temperatures therefore approach $T_\gamma$.  On the other hand, the effects at low redshifts are negligible, because the IGM is approaching isothermality anyway.

In summary, we do not expect the 21 cm background to be as useful for studying dark matter annihilation (as opposed to decay).  This is because a large fraction of the energy is injected at high redshifts, when the 21 cm background vanishes.  Fortunately, this is also precisely the regime to which the CMB is most sensitive, so together these techniques will offer a useful window into dark sector processes.

\section{\label{htwo} ${\rm H_{2}}$ formation and IGM chemistry}

Atomic species in gas of primordial composition lack low-lying energy levels. Thus, before the production of metals, molecular hydrogen is a critical coolant vital for star formation to proceed. An increased ${\rm H_{2}}$ abundance could also in principle change the temperature of the gas and affect 21 cm observations. Computing the production of ${\rm H_{2}}$ in the early universe has a long history stretching back to the first calculation by Saslaw \& Zipoy \cite{saslaw67}, but more recently it has been the subject of intense study as computational resources and estimates of reaction cross-sections have improved. Generally, the amount formed in pregalactic gas, $x_{\rm H_{2}} \sim 10^{-6}$ \cite{galli98,stancil98,lepp02,hirataH2_06}, is insufficient to be of physical consequence, although the abundance $x_{\rm H_{2}} \sim 10^{-4}$--$10^{-3}$ formed in the denser and hotter gas of collapsed halos (where reactions can proceed more quickly), {\it is} sufficient to trigger star formation \cite{abel02,bromm02}. If dark matter decay preheats and ionizes the IGM, this will boost ${\rm H_{2}}$ formation both due to the increased temperature (since reaction rates are highly temperature dependent) and the increased abundance of free electrons (which are a key catalyst). It would be exceedingly interesting if this boost was sufficient to affect subsequent structure formation, or left an observational signature. Recent estimates differ as to the magnitude of this boost \cite{padmanabhan05,biermann06,ripamonti06}. Here, we provide a more detailed look at ${\rm H_{2}}$ formation and consider some hitherto neglected effects.   

There are two main intermediaries by which ${\rm H_{2}}$ is produced in the gas phase: ${\rm H^{-}}$ and ${\rm H_{2}^{+}}$. In contrast to previous studies, Ref. \onlinecite{hirataH2_06} resolve all 423 rotational-vibrational levels of the ${\rm H_{2}^{+}}$ ion and show that the ${\rm H_{2}^{+}}$ pathway is greatly suppressed. This is because newly formed ${\rm H_{2}^{+}}$ ions are photo-dissociated by the CMB before they can decay to the ground state or undergo charge transfer to become ${\rm H_{2}}$ molecules. The key formation pathway is therefore through ${\rm H}^{-}$, via the reactions: (1) ${\rm H} + $e$^{-} \leftrightarrow {\rm H^{-}} + \gamma$, and (2) ${\rm H^{-}} + {\rm H} \leftrightarrow {\rm H_{2}} + $e$^{-}$. A net sink of ${\rm H^{-}}$ ions is mutual neutralization:  (3) ${\rm H}^{-} + {\rm H}^{+} \leftrightarrow 2{\rm H}$. The ${\rm H_{2}}$ production rate can be then obtained by assuming that the ${\rm H^{-}}$ ion takes its equilibrium value \cite{hirataH2_06}: 
\begin{equation}
\dot{n}_{\rm H_{2}}^{\rm form}= \frac{k_{1} k_{2} x_{e} x_{\rm HI}^{2} n^{2}}{k_{2}x_{\rm HI}n + k_{-1} + k_{3} x[{\rm H^{+}}] n}.
\label{eqn:H2_production}
\end{equation}
Here, $k_{1},k_{2},k_{3}$ refer to the forward reaction rates of reactions (1),(2),(3), $k_{-1}$ is the photo-detachment rate of ${\rm H}^{-}$ due to both the CMB and nonthermal spectral distortion photons from recombination, and $n$ is the total proper density of hydrogen nuclei. We use the fits for $k_{1},k_{2},k_{3}$ from \cite{stancil98} and calculate $k_{-1}$ as in \cite{hirataH2_06}. In particular, we take care to compute the non-thermal spectral distortion to the CMB from HI two-photon decay and Ly$\alpha$ photons from cosmological recombination (the two have roughly comparable contributions) by running RECFAST \cite{seager99} and performing the appropriate integrals \cite{switzer05,wong06}. We use the parametric fit to the two-photon profile given by \cite{nussbaumer84}. This component generally becomes important at $70< z < 120$, when CMB photons above the ${\rm H^{-}}$ photo-detachment threshold energy of $0.74$ eV lie in the steeply declining Wien tail, and before a significant fraction of distortion photons redshift below threshold. As we shall see, this is also the period of peak ${\rm H_{2}}$ production, even in cases where the IGM is pre-heated and ionized. Finally, we also include collisional ${\rm H_{2}}$ destruction processes, with reaction rates as given in \cite{oh02-h2}, generally the most significant being the charge exchange reaction ${\rm H_{2} + H^{+} \rightarrow H_{2}^{+} + H}$. While normally unimportant, they can become significant when the IGM is heated above $\sim 3000$ K.  

Before proceeding to calculate ${\rm H_{2}}$ abundances in our energy injection scenarios, we need to consider the importance of other sources of radiation. The energy injected into the IGM from dark matter decay/annihiliations introduces an additional non-thermal component to the radiation field, namely hydrogen Lyman series photons from atomic excitations (as well as two-photon decay products from ionized hydrogen atoms). Photodissociation of ${\rm H^{-}}$ and ${\rm H_{2}}$ could in principle retard ${\rm H_{2}}$ formation. In particular: (1) Ly$\alpha$ (and other) photons could photodissociate ${\rm H}^{-}$ (which has a $0.74$ eV threshold, and peaks at $1.4$ eV). However, this effect is strongly subdominant to the CMB, and (at $z < 120$) the spectral distortion from cosmological recombination at $z=1000$. The latter is easy to see: there must be at least one non-thermal photon per baryon from recombination. On the other hand, if $\chi_{i} \sim \chi_{e}$, then as a crude first approximation there are $\sim x_{e} \ll 1$ non-thermal photons per baryon from dark matter decay, much less than the contribution from recombination. (2) Lyman-Werner (LW) photons in the $11.2$--$13.6$ eV band could photo-dissociate ${\rm H_{2}}$ molecules. Unlike ${\rm H^{-}}$ photo-dissociation, this possibility only exists with high-redshift energy injection: the non-thermal distortion from recombination only extends up to $10.2$ eV, and rapidly redshifts to lower energies. On the other hand, Ly$\beta$ and higher series photons from atomic excitations can photo-dissociate ${\rm H_{2}}$. We can estimate this effect as follows. About $f_{\beta} \sim 15\%$ of excitation energy goes into Ly$\beta$ photons (or $\chi_{\beta} \sim f_{\beta} \chi_{e} \sim 5\%$ of injected energy), with a much smaller fraction going into higher order transitions (see \cite{pritchard06}). Each Ly$\beta$ photon scatters $n_{\rm scat} \sim 10$ times before being downgraded to a Ly$\alpha$ photon \cite{pritchard05}. Thus, Ly$\beta$ photons are produced at a rate $\dot{n}_{\beta}^{+}=m_{p}c^{2}n_{p} \chi_{\beta} \xi_{\rm X}/(h \nu_{\beta})$, and destroyed at a rate $\dot{n}_{\beta}^{-} = [(n_{p} \sigma_{\beta} c)/n_{\rm scat}] n_{\beta}$. Since the destruction timescale is much shorter than the Hubble time, we can consider the Ly$\beta$ photons to assume their time-independent value and obtain $n_{\beta}$ by setting $\dot{n}_{\beta}^{+}=\dot{n}_{\beta}^{-}$. The radiation field in the LW bands is then given by $J^{\rm LW} = hc/(4\pi) n_{\beta}$, or
\begin{eqnarray}
J_{21}^{\rm LW} & \approx & 10^{21} \frac{m_{p} c^{2} \chi_{\beta} \xi_{\rm X}}{4 \pi} \frac{n_{\rm scat}}{\nu_{\beta} \sigma_{\beta}} \\
& = & 2 \times 10^{-5} \xi_{-24} \left( \frac{\chi_{\rm B}}{0.05} \right) \left( \frac{n_{\rm scat}}{10} \right)  
\nonumber
\end{eqnarray}
where $J_{21}= J/(10^{-21} \, {\rm erg \, s^{-1} \, cm^{-2} \, sr^{-1} \, Hz^{-1}})$. Note there is no explicit redshift dependence, apart from possible dependence of $\xi_{\rm X}$ on redshift. Since $k_{\rm diss} = 1.6 \times 10^{-12} J_{21}^{\rm LW} \, {\rm s^{-1}}$ \cite{draine96}, this implies a dissociation rate per Hubble time of:
\begin{eqnarray}
f_{\rm diss} & = & \frac{k_{\rm diss}}{H(z)} \\
& = & 2.6 \times 10^{-2} \left( \frac{1+z}{100} \right)^{-1.5} \xi_{-24} \left( \frac{\chi_{\rm B}}{0.05} \right) \left( \frac{n_{\rm scat}}{10} \right).  \nonumber
\end{eqnarray}
Thus, at most a few percent of ${\rm H_{2}}$ molecules will be photo-dissociated by the ambient LW radiation field, and we will ignore this effect (of course, star formation seeded by ${\rm H_{2}}$ production could produce a much larger UV background with much more significant consequences for ${\rm H_{2}}$ chemistry). 

Could we observe the non-thermal spectral distortion to the CMB produced by Ly$\alpha$ photons from energy injection at high redshift? If $\chi_{e} \sim \chi_{i}$, then the comoving number density of Ly$\alpha$ photons is $n_{\gamma} \sim x_{e} n_{b}$. The observed intensity today from photons produced at $z_e$ is then
\begin{eqnarray}
\nu I_{\nu} & \sim & \frac{c}{4 \pi} n_{\gamma} \frac{E_{\gamma}}{(1+z_{e})} \\
& \sim & 7.8 \times 10^{-13} \left(\frac{x_{e}}{10^{-2}} \right) \left( \frac{1+z_{e}}{100} \right)^{-1} \, {\rm erg \, s^{-1} \, cm^{-2} \, sr^{-1}} \nonumber
\label{eq:inu}
\end{eqnarray}
at wavelengths $\lambda \sim 12 ([1+z_{e}]/100) \, \mu$m. By contrast, the observed extragalactic background light at these wavelengths is $\ga 10^{-6} \, {\rm erg \, s^{-1} \, cm^{-2} \, sr^{-1}}$ (see Fig. 9 of \cite{dole06}). Observing the Ly$\alpha$ and two-photon distortions to the CMB from cosmological recombination, which is brighter by a factor $\sim x_{e}^{-1} ([1+z_{e}]/1000) \sim 10$ but suffers from foregrounds of similar strength, is likewise still extremely challenging, although there have been proposals to do so \cite{fixsen02,wong06}. 

\begin{figure}
\resizebox{8cm}{!}{\includegraphics{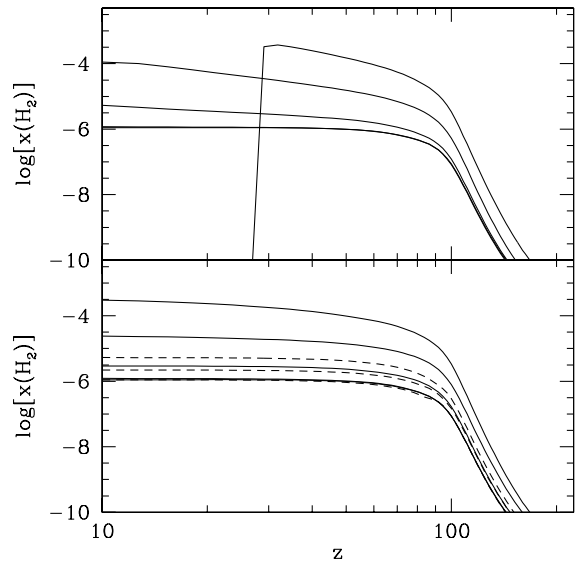}}
\caption{\label{fig:H2_redshift} ${\rm H_{2}}$ production in the IGM for different models of energy injection. Top panel: long-lived dark matter, with IGM histories as in Fig. \ref{fig:therm-thick}; curves take $\xi_{-24}=1,0.1,10^{-2},10^{-3}$, and $0$, from top to bottom. Note the rapid destruction of ${\rm H_{2}}$ in the top curve as the temperature climbs above 3000 K. Bottom panel: solid lines depict energy injection in transparency window, as in Fig. \ref{fig:therm-thin}; from top to bottom, curves take $\tau_{100}=1,0.1,10^{-2}$, and $10^{-3}$. Dashed curves are for dark matter annihilation, as in Fig. \ref{fig:therm-ann}, with $\langle \sigma v\rangle = 25,5.6$, and $1.25\times 10^{-26} \, {\rm cm^{3} \, s^{-1}}$ from top to bottom.}
\end{figure}

Since these contributions to the radiation field are negligible, ${\rm H_{2}}$ production in the case of an IGM heated/ionized by dark matter decays/annihilation can then be obtained by straightforwardly integrating equation (\ref{eqn:H2_production}) with the previously calculated temperature and ionization histories. The results are shown in Fig. \ref{fig:H2_redshift}. We see that the boost in ${\rm H_{2}}$ production can be substantial in dark matter decay scenarios, with values approaching $x_{\rm H_{2}} \sim 5 \times 10^{-4}$, which is comparable to the maximal asymptotic abundance of $x_{\rm H_{2}} \sim 10^{-3}$ in gas cooling via atomic transitions from $T > 10^{4}$ K (the ``maximally heated" case of long-lived dark matter with $\xi_{-24}=1$ tends to approach this value, but ${\rm H_{2}}$ is rapidly destroyed by charge exchange reactions as the IGM temperature climbs past 3000 K). The latter ``freeze-out" value can be understood from simple timescale arguments \cite{oh02-h2}. Note that we have only computed ${\rm H_{2}}$ formation at the mean density of the IGM---it will be more efficient and rapid in the higher temperature and denser environments of collapsed halos. Thus, close to the maximal amount of ${\rm H_{2}}$ can be formed, certainly comparable to the abundance $x_{\rm H_{2}} \sim 10^{-4}$ needed for molecular hydrogen to trigger cooling and star formation in collapsed halos (e.g., see \cite{tegmark97}). In particular, it exceeds the ${\rm H_{2}}$ abundance formed in low mass halos (where reactions are slower due to lower temperatures) catalysed by the low primordial electron fraction $x_{e}\sim 2 \times 10^{-4}$. Thus, some of these dark matter decay scenarios may be ruled out on the grounds that they would naturally seed a good deal of high redshift star formation which would violate bounds on the observed \emph{WMAP} optical depth. However, a variety of interlocking feedback mechanisms are at play, so it is difficult to make quantitative claims without further study. 

For example, one possible caveat to the claim that early preheating/reionization would seed early star formation is that the entropy of the IGM will suppress gas accretion onto halos and instead exert a negative feedback effect \cite{oh03-entropy}. However, early reionization by decaying dark matter differs in one crucial respect from early reionization by stars, the scenario envisaged by Ref.~\onlinecite{oh03-entropy}: unlike early star formation, reionization by decaying dark matter is accompanied by a negligible LW background. Ref. \onlinecite{oh03-entropy} showed that the cores of gas accreted from a high entropy IGM had low enough densities that a small LW background would suffice to destroy any ${\rm H_{2}}$ formed---i.e., the photo-dissociation timescale was much shorter than the ${\rm H_{2}}$ cooling time. By contrast, in our present scenario the LW background is negligible and indeed, ${\rm H_{2}}$ can form and survive at the mean density of the IGM. Thus, as long as the gas can contract to sufficient density that the cooling time falls below the expansion timescale, the effect of the entropy floor is unimportant. Of course, once some star formation takes place, the LW background rises and ${\rm H_{2}}$ destruction in low density cores and the IGM proceeds. Therefore, detailed study is necessary to understand if an early epoch of preheating and copious ${\rm H_{2}}$ production will indeed result in extensive star formation in violation of WMAP optical depth bounds. 

There are a few general features worth noting about ${\rm H_{2}}$ production in these scenarios. Most of the ${\rm H_{2}}$ in all scenarios is made at $z\sim 100$; this is the highest redshift at which ${\rm H^{-}}$ photo-detachment from the high-energy tail of CMB becomes unimportant, yet where the IGM gas is still sufficiently dense that reactions can proceed rapidly. This era of peak ${\rm H_{2}}$ production is fairly independent of temperature or ionization history in the different scenarios. The boost in ${\rm H_{2}}$ production is primarily due to the increased free electron fraction; for low $x_{e}$, we see from equation (\ref{eqn:H2_production}) that $x_{\rm H_{2}} \propto x_{e}$ (indeed, the peak ${\rm H_{2}}$ abundance can roughly be estimated from $x_{\rm H_{2}} \sim k_{m}(z_{f}) n(z_{f}) x_{e}(z_{f}) t_{\rm H} (z_{f})$, where $k_{m} \approx k_{1} k_{2}/k_{-1}$, and $z_{f} \sim 100$ is the redshift at which $k_{m}$ peaks). The ${\rm H_{2}}$ formed should not have a significant effect on the temperature of the IGM: since the ${\rm H_{2}}$ cooling function $\Lambda_{\rm H_{2}} \propto T^{4}$ for $T < 3000$ K, the cooling time is:
\begin{equation}
t_{\rm cool} = 9 \times 10^{8} \left( \frac{1+z}{100} \right)^{-3} \delta \left( \frac{x_{\rm H_{2}}}{10^{-4}} \right)^{-1} \left( \frac{T}{1000 \, {\rm K}} \right)^{-3} \yr
\end{equation}
where $\delta$ is the gas overdensity. This is substantially greater than the Hubble time and the Compton cooling time $t_{C} = 1.2 \times 10^{6} ([1+z]/100)^{-4} (x_{e}/10^{-2})^{-1} \yr$. The effects of ${\rm H_{2}}$ cooling are only important in dense virialized halos. 

Apart from the possible effect of seeding high-redshift star formation, there are few observable consequences of this large amount of early ${\rm H_{2}}$ formation. They could potentially increase fluctuations in Ly$\alpha$ coupling (due to the consumption of LW photons in photo-dissociation regions), but this is likely difficult to detect.       

\section{\label{disc}Discussion}

The 21 cm transition is, at least in principle, a window into the dark ages of structure formation at $z \ga 50$.  We have argued that, because of the overall simplicity of the (known) physics at that time -- the expanding Universe, hydrogen recombination, and linear gravitational growth -- it presents a unique probe of both cosmology \cite{loeb04, bharadwaj04-vel} and exotic processes such as dark matter decay and annihilation.  The heating and ionization induced by the decay (or annihilation) products can significantly affect the IGM.  These processes can modify the CMB power spectrum \cite{chen04-decay, pierpaoli04, mapelli06}, but this is a relatively insensitive measure because it requires a large $\bxion$ for scattering to be significant.

Here we have shown that the 21 cm history is a sensitive measure of decay and annihilation during the dark ages (see also \cite{shchekinov06}), because it directly measures the thermal history of the IGM.  In the standard calculation, adiabatic cooling drives $\bar{T}_K$ to such low levels that heating the IGM significantly requires much less energy than ionizing it.  We have shown that dark matter with lifetimes $\sim 10^{24}$--$10^{27} \sec$ can substantially affect the 21 cm background (see Figs.~\ref{fig:signal-thick} and \ref{fig:signal-thin}).  These timescales are about three to four orders of magnitude longer than those probed by the CMB.  The improvement is considerably smaller for annihilation scenarios, because those tend to inject most of their energy nearer the surface of last scattering, when the CMB is more sensitive and the 21 cm background vanishes.

We have made predictions for both the sky-averaged signal -- which measures the total energy deposition rate \cite{shchekinov06} -- and the 21 cm power spectrum.  Because the temperature, ionization fraction, and \lya background all affect the 21 cm signal,  the overall amplitude and redshift evolution (at \emph{any} scale) of $P_{21}$ can provide powerful constraints on such exotic processes:  reasonable scenarios produce order unity effects on the power spectrum and also introduce non-trivial redshift dependence if the IGM ever becomes hotter than the CMB.  Thus \emph{any} measurement of $P_{21}$ during the dark ages (such as those advocated by \cite{loeb04}) will be useful in this context, even if it only measures fluctuations on rather large scales.

More detailed measurements can begin to constrain the decay and annihilation processes themselves by inferring the properties of the products that interact with the IGM.  For example, if the decay produces either soft ($\la 3 \keV$) or extremely hard ($\ga 10 \GeV$) photons, the energy will be deposited into the IGM nearly instantaneously \cite{chen04-decay}.  In this case the effects will be most obvious at lower redshifts, when there is more time for heating to take place and Compton coupling with the (spatially uniform) CMB is weaker.  On the other hand, if the IGM is optically thin to the products (or if the dark matter annihilates), the consequences are more confined to higher redshifts (where the optical depth is larger because of the increased density).  More subtly, the interaction processes between the IGM and the decay products determine the fluctuations in the ionizing and heating rates, which in turn affect $P_{21}$ (see Figs.~\ref{fig:breakdown} and \ref{fig:signal-ann}).  In some cases, this could even introduce extra spatial dependence into the power spectrum, although we have not examined such effects here.

Rather than examining specific particle physics models, we used a generic and flexible formulation for decay and annihilation.  We refer the interested reader to the discussions in \S \ref{particles} and in Refs.~\onlinecite{mapelli06, ripamonti06} for the thermal and ionization histories in specific models.  In general, however, we note that among recently popular models, decaying light dark matter (such as axinos and sterile neutrinos) would have the strongest effects.  The decay of heavy dark matter could also affect the thermal history, though this depends strongly on the allowed decay channels.  Neutralino annihilation, or the annihilation of light dark matter, could also provide an interesting signal.

We have also shown that the heat deposited in the IGM, as well as the excess ionization, can affect the chemistry of the IGM.  In particular, the increased temperature and ionization can dramatically increase the rate of H$_2$ formation.  While the resulting abundance can be up to orders of magnitude higher than the standard value (and in some cases comparable to the maximal asymptotic abundance $x_{\rm H_{2}} \sim 10^{-3}$ for gas phase ${\rm H_{2}}$ formation), it is not an important coolant at the low densities of the IGM.  Thus its direct observable effects are small. However, the increased ${\rm H_{2}}$ abundance will strongly simulate early star formation in dense halos, quite possibly violating WMAP constraints on $\tau_{e}$. With better modeling, observations of the first stars and $\tau_{e}$ may also provide limits on dark matter decay and annihilation.

We must of course acknowledge the tremendous difficulty posed by 21 cm observations at $z \ga 50$, the regime in which dark matter decay signatures would be cleanest.  The principal challenge is the enormous brightness of the Galactic synchrotron foreground, which has a brightness temperature $T_{\rm sky} \ga 10^4 \kel$ at the relevant frequencies of $\sim 30 \MHz$.  It will make measurements of the smoothly varying $\bdtb$ extremely difficult; searches for fluctuations will probably be much easier (though still well beyond current capabilities).  The largest transverse wavenumber observable by an array distributed in a circle with radius $R_{\rm max}$ is $k_{\perp,{\rm max}} \approx 0.2 \, [50/(1+z)] \, (R_{\rm max}/2 \, {\rm km}) \Mpcinv$.  For a crude sensitivity estimate, we consider an array with uniform baseline coverage observing the spherically-averaged signal.  Then the error on the power spectrum at wavenumber $k$ would be \cite{mcquinn05-param}
\begin{eqnarray}
\sqrt{\frac{k^3 \delta P_{21}}{2 \pi^2}} & \sim & \frac{0.1 \mkel}{\epsilon^{1/4} \, f_{\rm cov}} \, \left( \frac{k}{0.04 \Mpcinv} \right)^{3/4} \left(\frac{T_{\rm sky}}{10^4 \kel} \, \frac{2 \, {\rm km}}{R_{\rm max}} \right) \nonumber \\
& & \left( \frac{10 \MHz}{B} \right)^{1/4} \left( \frac{1000 \hr}{t_{\rm int}} \right)^{1/2} \left( \frac{1+z}{50} \right),
\label{eq:sens}
\end{eqnarray}
where $f_{\rm cov} \equiv A_e/(\pi R_{\rm max}^2)$ is the array covering factor,  $A_e$ is its effective area, $B$ is the bandwidth of the observation, $t_{\rm int}$ is the total integration time, and we have binned the data in segments of logarithmic length $\epsilon k$.  Here we have assumed that $k$ is much larger than the wavenumber corresponding to the total bandwidth of the experiment ($\sim 0.015 \Mpcinv$ for $B=10 \MHz$) and much smaller than $k_{\perp,{\rm max}}$.  Most importantly, we have assumed a large field of view -- of order one steradian -- corresponding to baselines of order 10 m; thus, the instrument must be composed of many small antennae.  Our parameter choices in equation~(\ref{eq:sens}) allow a direct comparison with the figures in this paper at $z \sim 50$ if $R_{\rm max}=2$ km.  Clearly, several square kilometers of collecting area are required to produce any useful limits.  Moving to larger scales can help slightly, but foreground removal will probably compromise measurements at $k \la 0.01 \Mpcinv$ (see \S 9.3 of Ref.~\onlinecite{furl06-review} and references therein).  Fortunately, more compact array designs improve the sensitivity over a uniform baseline distribution by factors of a few (see Fig.~6 of Ref.~\onlinecite{mcquinn05-param}).  Exploration of the highly-redshifted 21 cm sky is just beginning, and over the next few years we should learn much more about what is possible.  When experiments to open up the dark ages do eventually come along, they will provide important constraints on dark matter decay and annihilation -- which can plausibly have order unity effects on the 21 cm signals.  

\section{acknowledgments}

S.R.F. thanks the Tapir group at Caltech for their hospitality while much of this work was completed and M. McQuinn for helpful discussions.  E.P. is an ADVANCE fellow (NSF grant AST-0340648) and is also supported by NASA grant NAG5-11489. S.P.O gratefully acknowledges NSF grant AST-0407084 and NASA grant NNG06GH95G for support. 


\end{document}